\documentclass[fleqn,usenatbib]{mnras}

\usepackage{newtxtext,newtxmath}
\usepackage[T1]{fontenc}

\DeclareRobustCommand{\VAN}[3]{#2}
\let\VANthebibliography\thebibliography
\def\thebibliography{\DeclareRobustCommand{\VAN}[3]{##3}\VANthebibliography}

\usepackage{graphicx}	% Including figure files
\usepackage{amsmath}	% Advanced maths commands

\usepackage{siunitx}

\usepackage{caption}
\usepackage{subcaption}
\usepackage{multicol}

\usepackage{hyperref}
\usepackage[hyphenbreaks]{breakurl}

\DeclareSIUnit \parsec {pc}
\DeclareSIUnit \mas {mas}
\DeclareSIUnit \arcsec {as}
\DeclareSIUnit \photon {ph}

\title{Behind the Mask: can HARMONI@ELT detect biosignatures in the reflected light of Proxima b?}

\author[S. R. Vaughan et al.]{Sophia R. Vaughan$^{1}$, %\thanks{E-mail: mn@ras.org.uk (KTS)}
Jayne L. Birkby$^{1}$,
Niranjan Thatte$^{1}$,
Alexis Carlotti$^{2}$,
Mathis Houll\'{e}$^{3}$,
\newauthor
Miguel Pereira-Santaella$^{4}$,
Fraser Clarke$^{1}$,
Arthur Vigan$^{5}$,
Zifan Lin$^{6}$,
and Lisa Kaltenegger$^{7,8}$
\\
$^{1}$Department of Physics, University of Oxford, Oxford, OX1 3RH, UK\\
$^{2}$Univ. Grenoble Alpes, CNRS, IPAG, 38000 Grenoble, France\\
$^{3}$Université Côte d'Azur, Observatoire de la Côte d'Azur, CNRS, Laboratoire Lagrange, 06304 Nice, France\\
$^{4}$Instituto de F\'isica Fundamental, CSIC, Serrano 123, 28006 Madrid, Spain\\
$^{5}$Aix Marseille Univ., CNRS, CNES, LAM, 13388 Marseille, France\\
$^{6}$ Department of Earth, Atmospheric, and Planetary Sciences, Massachusetts Institute of Technology, 77 Massachusetts Avenue, Cambridge, MA 02139, USA\\
$^{7}$ Carl Sagan Institute, Cornell University, 302 Space Sciences Building, Ithaca, NY 14853, USA\\
$^{8}$ Astronomy Department, Cornell University, 302 Space Sciences Building, Ithaca, NY 14853, USA\\
}
\date{Accepted 2024 January 17. Received 2024 January 11; in original form 2023 November 20}
\pubyear{2024}

\begin{document}
\label{firstpage}
\pagerange{\pageref{firstpage}--\pageref{lastpage}}
\maketitle

\begin{abstract} % 247/250 words
%Context
Proxima b is a rocky exoplanet in the habitable zone of the nearest star system and a key test case in the search for extraterrestrial life.
%Aims
Here, we investigate the characterization of a potential Earth-like atmosphere around Proxima b in reflected light via molecule mapping, combining high resolution spectroscopy (HRS) and high contrast imaging, using the first-generation integral field spectrograph HARMONI on the $39$-m Extremely Large Telescope.
%Methods
We simulate comprehensive observations of Proxima b at an assumed $45^{\circ}$ inclination using HARMONI’s High Contrast Adaptive Optics mode, with spatial resolution $\sim\SI{8}{\mas}$ ($\SI{3.88}{\mas}$/spaxel) and spectral resolving power $R\simeq17,000$ between $1.538$--$1.678\SI{}{\micro\meter}$, containing the spectral features of water, carbon dioxide and methane. Tellurics, stellar features, and additional noise sources are included, and removed using established molecule mapping techniques.
%Results
We find that HARMONI’s current focal plane mask (FPM) is too large and obscures the orbit of Proxima b and thus explore smaller and offset FPMs to yield a detection. A $\rm{S/N}=5$ detection of Proxima b’s reflected light, suitable for atmospheric characterisation, is possible with such modifications, requiring a minimum of $20$ hours, but ideally at least $30$ hours of integration time. We highlight that such detections do not scale with the photon noise, hence suitably detailed simulations of future instruments for the ELTs are needed to fully understand their ability to perform HRS observations of exoplanet atmospheres.
%Conclusions
Alterations to the HARMONI FPM design are feasible at this stage, but must be considered in context of other science cases.
\end{abstract}


%%%%%%%%%%%%%%%%%%%%%%%%%%%%%%%%%%%%%%%%%%%%%%%%%%

%%%%%%%%%%%%%%%%% BODY OF PAPER %%%%%%%%%%%%%%%%%%

\section{Introduction}
\label{sec:intro}

One of the key goals of exoplanet science is the atmospheric characterisation of Earth-sized planets in the habitable zones of Sun-like stars. An important focus of these studies will be the search for biosignatures; indicators of life, such as the combination of oxygen (O$_2$), carbon dioxide (CO$_2$), water (H$_2$O) and methane (CH$_4$) \citep{Meadows2018}. However, current and near future planned observatories will find it very challenging to characterise a temperate rocky exoplanet in the habitable zone of a Sun-like star, due to unfavourable ratios between the planet-star radii and contrast, as well as very close spatial separations and low transit probabilities. Studies have thus turned to smaller stars, i.e. M-dwarfs, where a reduced stellar radius and luminosity leads to closer-in conventional habitable zones and therefore more favourable ratios and transit probabilities giving greater observability \citep{Charbonneau2007}. In addition, M-dwarfs are the most common type of star in the galaxy with approximately $250$ within $\SI{10}{\parsec}$ of the Sun \citep{Reyle2021, Reyle2023}. These stars also have a higher occurrence rate of rocky planets than FGK stars \citep{Mulders2015}, meaning that the prevalence of habitable worlds may be significantly influenced by the environment of M-dwarf host stars \citep{Shields2016}. The habitability of an M-dwarf Earth is currently an open question due to its dependence on a wide array of factors such as atmospheric loss \citep[e.g.][]{Khodachenko2007}, tidal locking \citep[e.g.][]{Showman2013} and photosynthetic viability \citep[e.g.][]{Kiang2007, Claudi2021}. Of key importance is determining if these worlds have atmospheres. M-dwarfs can be very active stars \citep{Vida2019, Yang2017} and it is possible that the stellar activity has eroded the planetary atmosphere which may impact abiogenesis \citep{Khodachenko2007, Kreidberg2019} but observations with JWST of a handful of transiting M-dwarf systems should give a first indication of whether atmospheres are retained in the conventional habitable zone \citep[e.g. LHS 475\,b, TRAPPIST-1\,b and TRAPPIST-1\,c][]{Lustig-Yaeger2023, Greene2023, Zieba2023}. If these planets have atmospheres, it is likely that they contain CO$_2$ due to the balance between volcanic and tectonic outgassing of CO$_2$ and weathering sinks \cite[e.g.][]{Foley2018}. Therefore CO$_2$ is a good tracer of the presence of a rocky planet atmosphere, as well as being a potential biosignature in combination with other species.

Proxima b, the nearest exoplanet ($\SI{1.3}{\parsec}$), offers the opportunity for a detailed close-up study of the environment of a rocky exoplanet in the habitable zone of a bright M-dwarf \citep{Anglada2016, Jenkins2019, Damasso2020, Faria2022}. An exoplanet's atmosphere can be characterised through its transmission spectrum, thermal emission spectrum or reflection spectrum. The Extremely Large Telescopes (ELTs) could detect CO$_2$ and CH$_4$ on transiting planets, but O$_2$ and H$_2$O may require an unfeasible number of transits to detect \citep{Currie2023, Hardegree-Ullman2023}. However, like many nearby M-dwarfs planets, Proxima b does not transit \citep[][and references therein]{Gilbert2021}. Therefore, this world must be characterised through its reflection or thermal emission spectra. While a full thermal phase curve with JWST may potentially distinguish between bare rock and an atmospheric presence, significant observing times of months are needed for molecular detection \citep{Kreidberg2016}. Furthermore, the future planned Habitable Worlds Observatory does not have Proxima b in its current target list \footnote{See NASA Exoplanet Exploration Program's Mission Star List for the Habitable Worlds Observatory. Available online: \url{https://exoplanets.nasa.gov/internal_resources/2645_NASA_ExEP_Target_List_HWO_Documentation_2023.pdf}} which will directly image the Habitable Zones of other nearby stars. A promising alternative is to leverage the better spatial resolution of the ground-based ELTs, in combination with high spectral resolution and high contrast imaging, to resolve the planet from its host star and obtain its thermal and reflection spectrum directly using the technique of molecule mapping \citep[e.g.][]{SI2015,HH2018-2, PddlRD2018, Wang2018, CG2021, Petrus2021, Wang2021, Ruffio2021, Ruffio2023}.

Molecule mapping works by leveraging the spatial resolution obtainable with a large mirror using adaptive optics (AO) to suppress contamination by diffracted star light at the planet's location. An integral field or long slit spectrograph then takes a spectrum of each `spaxel' (spatial pixel) in an image including at the planet's location. Since most spaxels will not contain the planet's spectrum, it is possible to create a data driven model of the stellar contamination on the exoplanet's spectrum. This technique is aided by high spectral resolution which separates the planet's spectral lines from those of the contamination. However, this does not remove the photon noise caused by the stellar spectrum so in most cases the planet's spectrum will be very low signal-to-noise. Nonetheless, information can be obtained by cross correlating each spaxel in the image with a model of the exoplanet's spectrum. This combines the signal of the planet's spectrum across wavelength resulting in a higher cross correlation value than the background noise. This technique can be used to obtain the low resolution albedo function of the planet using a method similar to e.g. \citet{Martins2018} and high resolution information through retrieval methods which test many planet models to put constraints on planetary properties \citep[e.g.][]{Ruffio2019, Hoch2020, Ruffio2021, Zhang2021, Patapis2022, Wang2022, Xuan2022, Landman2023, Xuan2023}.

There are many AO-enabled Integral Field Spectrographs (IFSs) among the next generation of instruments for the ELTs. These include HARMONI, METIS, ANDES and PCS on the ELT, GMTIFS and GMagAO-X + IFS on the Giant Magellan Telescope (GMT) and IRIS and MICHI on the Thirty Meter Telescope (TMT) to name a few \citep{HARMONI,METIS,ANDES,PCS,GMTIFS,GMagAOX,IRIS,MICHI}. As molecule mapping is a photon-limited technique, it typically makes use of these ground-based instruments at wavelengths shorter than $\approx \SI{5}{\micro\meter}$ due to the thermal contamination from room temperature telescope optics and the Earth's atmosphere \citep{SI2015}. HARMONI and METIS on the ELT form a powerful first light pair for molecule mapping. This is because HARMONI operates at wavelengths less than $\SI{2.5}{\micro\meter}$ where reflected light dominates the spectrum of a temperate rocky exoplanet \citep[see Fig. 7 in][]{Turnbull2006} while the high resolution part of METIS will target mostly thermal emission between $\SI{3}{}$--$\SI{5}{\micro\meter}$. In addition, RISTRETTO@VLT will operate between $\SI{0.62}{}$--$\SI{0.84}{\micro\meter}$, aiming at detecting the oxygen A-band, $\SI{0.759}{}$--$\SI{0.771}{\micro\meter}$, on Proxima b \citep{RISTRETTOold,RISTRETTO}. Combined they have the potential to give a holistic view of a planet's atmospheric properties, including its energy budget, and robust measurements of its atmospheric constituents including the four key biomarkers O$_2$, CO$_2$, H$_2$O and CH$_4$. \citet{SI2015} studied the potential of METIS to characterise the thermal properties of the atmosphere of a planet similar to Proxima b. In this work we focus on the complementary reflection spectrum, by simulating observations for HARMONI. Previous work \citep{Houlle2021, Bidot2023}, has demonstrated HARMONI's ability to detect the thermal emission of young, widely-separated gas giants using the molecule mapping technique. Following the work of e.g. \citet{Wang2017}, \citet{Houlle2021} and \citet{Patapis2022}, we robustly estimate HARMONI's potential to characterise Proxima b including the effects of Earth's atmosphere, the optics of the ELT and HARMONI's, detector performance and several noise sources as in \citet{Houlle2021} (described fully in Section \ref{sec:methods}). Additionally, due to the small on sky separation and short orbital period ($\approx\SI{11.2}{\day}$) of Proxima b, its on sky position and velocity can change appreciably during an observation. This motion is an effect currently unique to this system and is accounted for in the simulations. 

This paper is laid out as follows. Section \ref{sec:methods} describes the simulation of HARMONI observations for spatially resolved exoplanet systems made with the High Contrast Adaptive Optics (HCAO) mode. Section \ref{sec:molecule mapping} demonstrates how the molecule mapping technique can be used to recover the signal of a fiducial exoplanet's reflected light while Section \ref{sec:results} demonstrates how HARMONI can be used to study Proxima b specifically. We discuss here also the possibility of relatively minor modifications to the HARMONI instrument design to best enable this. Section \ref{sec:discussion} discusses the arguments regarding the change to the instrument design. We conclude in Section \ref{sec:conclusions}.

%%%%%%%%%%%%%%%%%%%%%%%%%%%%%%%%%%%%%%%%%%%%%%%%%%
%%%%%%%%%%%%%%%%%%%%%%%%%%%%%%%%%%%%%%%%%%%%%%%%%%
%%%%%%%%%%%%%%%%%%%%%%%%%%%%%%%%%%%%%%%%%%%%%%%%%%

%%%%%%%%%%%%%%%%%%%%%%%%%%%%%%%%%%%%%%%%%%%%%%%%%%
\section{Simulating HARMONI}
\label{sec:methods}

HARMONI is an IFS that will be one of the first instruments mounted on the ELT \citep[]{Tecza2009,Thatte2010,Thatte2014,Thatte2016,Thatte2020,Thatte2021,Thatte2022}. It is a versatile instrument with: a non-simultaneous wavelength coverage between $\SI{0.47}{}$ -- $\SI{2.45}{\micro\meter}$; a choice of three spectral resolutions; four spatial resolutions and several AO modes including a high contrast mode. For spatially resolved observations of exoplanets, the HCAO mode \citep[]{Carlotti2018, Houlle2021} facilitates the required high-contrast observations with coverage of the H and K bands between $\SI{1.45}{}$ -- $\SI{2.45}{\micro\meter}$ at a range of resolutions. This work simulates observations with this mode using the H-high grating ($\SI{1.538}{}$ -- $\SI{1.678}{\micro\meter}$, $R=17385$) with a spatial sampling of $\SI{3.88}{\mas}$ which oversamples the spatial resolving power.

The HCAO mode can be configured with one of two available apodizers (named SP1 and SP2) and one of three partially transmissive focal plane masks (FPMs). The apodizers reside in the pupil plane and modify the point spread function (PSF) of the instrument to create a dark annulus where diffraction is suppressed around the central peak. The FPMs resides in the focal plane mask wheel and reduces the flux of the PSF core by a factor of $10^4$ which allows longer integration times to be used for bright stars. HARMONI uses an atmospheric dispersion corrector (ADC) optimised for a single airmass meaning there will be residual dispersion. Therefore, each of the FPM is elongated in one direction to better match the shape of PSF core. In this work we use only the SP1 apodizer with the H band (smallest) focal plane mask.

We simulate observations for HARMONI's HCAO mode using a similar method to that in HSIM v310\footnote{https://github.com/HARMONI-ELT/HSIM} \citep[]{Zieleniewski2015} but with modifications for multi-exposure simulations to reduce computation time and output file size. Our simulations also allow for the instrumental parameters to be modified to simulate changes to HARMONI's design and different IFSs. 
%%%%%%%%%%%%%%%%%%%%%%%%%%%%%%%%%%%%%%%%%%%%%%%%%%
%%%%%%%%%%%%%%%%%%%%%%%%%%%%%%%%%%%%%%%%%%%%%%%%%%

%%%%%%%%%%%%%%%%%%%%%%%%%%%%%%%%%%%%%%%%%%%%%%%%%%

\subsection{Modeling the Orbits}
\label{sec:orbit}
Due to Proxima b's short orbital period and HARMONI's small spatial sampling, the planet may not remain on the same spaxel throughout a night of observations. In HCAO mode, the instrument rotator will track the parallactic angle, otherwise known as pupil tracking, to keep atmospheric dispersion at a fixed angle. This means the position angle of the image will appear to rotate during an observing run. Over a night ($\sim10$ hours), this rotation corresponds to $\sim\SI{150}{\degree}$ and is accounted for in our simulations. A smaller but important effect is the change in Proxima b's position due to its relatively short orbital period and proximity to the Solar System. For a face on orbit the position of the planet will rotate by $\sim\SI{13}{\degree}$ ($\sim2$ spaxels at the separation of Proxima b) over $10$ hours.

As we aim to be as realistic as possible with the orientation of the Proxima system, we include the orbits of both the star and planet in the simulation although the stars motion is insignificant. Due to the spatial and spectral resolution of the simulation, it is sufficiently accurate to model the orbits as ellipses. For Proxima b's orbit we use measured parameters from radial velocity studies (see Table \ref{tab:proxima params}) and Proxima Centauri we assume reflex motion due to Proxima b only. The orientation of the orbits are specified by three parameters:
\begin{enumerate}
    \item Inclination, $i$: the angle between the plane of the sky and the plane of the orbit.
    \item The longitude of the ascending node, $\Omega$: a rotation in the plane of the sky.
    \item The argument of periastron, $\omega$: a rotation in the plane of the orbit.
\end{enumerate}

Unfortunately, the longitude of the ascending node is currently unknown and the inclination is only weakly constrained by the non-detection of transits \citep[and references therein]{Gilbert2021}. We choose to set the eccentricity to $0$, close to the value measured by \citet{Faria2022}, which means the argument of periastron no longer affects the shape of the orbit. An intermediate inclination of $\SI{45}{\degree}$ and a longitude of the ascending node of $\SI{90}{\degree}$ are chosen for Proxima b's orbit in this work. This choice does not affect the maximum elongation of Proxima b but does affect the duration of and Doppler shift at maximum elongation. Section \ref{sec:unknownorbit} discusses the difficulty of determining Proxima b's full 3-D orbit prior to these observations and how this analysis might proceed if the orbit is unknown. The orbit chosen represents a good compromise between having large radial velocity shifts and spending longer at maximum elongation. The mean longitude and accompanying reference time are used to define the phase of the orbit at a given time. 

Our treatment of Proxima b's orbit differs from that of \citet{SI2015} in their simulations of METIS as their planet stays at quadrature with a Doppler shift of $\SI{30}{\kilo\meter\per\second}$ throughout each observation. Additionally, their planet is also slightly larger ($R_p=1.5R_\oplus$) and closer to Proxima Centauri ($a=\SI{0.032}{\astronomicalunit}$) than Proxima b.

%%%%%%%%%%%%%%%%%%%%%%%%%%%%%%%%%%%%%%%%%%%%%%%%%%

\begin{table}
\begin{center}   
    \begin{tabular}{m{0.52\columnwidth}m{0.28\columnwidth}m{0.1\columnwidth}}
        \hline
        Spectral Parameters & Value & Ref \\
        \hline
        Stellar Type & M5.5 V & $1$ \\
        Right Ascension & $14:29:42.94613$ & $2$ \\
        Declination & $-62:40:46.16468$ & $2$ \\
        Distance & $\SI{1.302}{\parsec}$ & $2$ \\
        PHOENIX Model & $T=\SI{3000}{\kelvin}$ & $3$ \\
         & $\log(g)=5$ \newline $[Fe/H]=0$ \newline $[\alpha/M]=0$ & $3$ \newline $3$ \newline $3$ \\
        Stellar Rotation Period, $P_{rot,\star}$ & $\sim\SI{90}{\day}$ & $4,5$ \\
        Magnitude \textit{(H)} & $4.8$ mag & $6$ \\
        Proxima b Spectral Model & 1 bar oxic atmosphere & $7$ \\
        Proxima b Geometric Albedo (average) & \textbf{0.23} & $7$ \\
        Proxima b Minimum Mass, $M_p \sin i$ & $1.27 M_\oplus$ & $5$ \\
        Proxima b Radius (estimate), $R_p$ & $1.07 R_\oplus$ & $8$ \\   
        \hline
        Orbital Parameters & Value & Ref \\
        \hline
        Systemic Velocity, $V_{sys}$ & $\SI{-22.204}{\kilo\meter\per\second}$ & $9$ \\
        Radial Velocity Amplitude, $K_\star$ & $\SI{1.24}{\meter\per\second}$ & $4$ \\
        Semi-major axis, $a$ & $\SI{0.0485}{\astronomicalunit}$ & $4,5$ \\
        Orbital Period, $P$ & $\SI{11.186}{\day}$ & $4,5,10,11$ \\
        Orbital Velocity, $K_p$ & $\SI{47.2}{\kilo\meter\per\second}$ & * \\
        Eccentricity, $e$ & $0$ & * \\ 
        Inclination, $i$ & $\SI{45}{\degree}$ & * \\
        Longitude of the Ascending Node, $\Omega$ & $\SI{90}{\degree}$ & * \\
        Mean Longitude, $\lambda$ & $\SI{110}{\degree}$ & $5$ \\
        Reference time for $\lambda$ (JD), $T_\lambda$ & $2451634.73146$ & $5$ \\
    \end{tabular}
\caption{Parameters of the Proxima Centauri system used to simulate Proxima b's orbit and information on the spectra used in this work. \newline
        $1$ \citet{Bessell1991};
        $2$ \citet{GAIADR3};
        $3$ \citet{Husser2013};
        $4$ \citet{Faria2022};
        $5$ \citet{Anglada2016};
        $6$ \citet{Cutri2003};
        $7$ \citet{Lin2020};
        $8$ \citet{Bixel2017};
        $9$ \citet{Kervella2017};
        $10$ \citet{Jenkins2019};
        $11$ \citet{Damasso2020};
        * assumed  
        }
\label{tab:proxima params}
\end{center}
\end{table}

%%%%%%%%%%%%%%%%%%%%%%%%%%%%%%%%%%%%%%%%%%%%%%%%%%
\subsection{Spectrum of Proxima b}
\label{sec:spectra}

HARMONI will be sensitive to the reflected light of Proxima b meaning the planet's spectrum, $F_p$, is a Doppler shifted and rotationally broadened copy of the stellar spectrum, $F_s$, \citep{Spring2022} modulated by the planets geometric albedo, $A_g$, as shown in Equation \ref{equ:fp}. 

\begin{equation}
    \label{equ:fp}
    F_p(\lambda, v_{Dop}) = F_s(\lambda, v_{Dop}) \times A_g \times g(\alpha) \times (R_p/a)^2
\end{equation}

\begin{equation}
    \label{equ:ga}
    g(\alpha) = \frac{1+\cos(\alpha)}{2}
\end{equation}

\begin{figure}
	\includegraphics[width=\columnwidth]{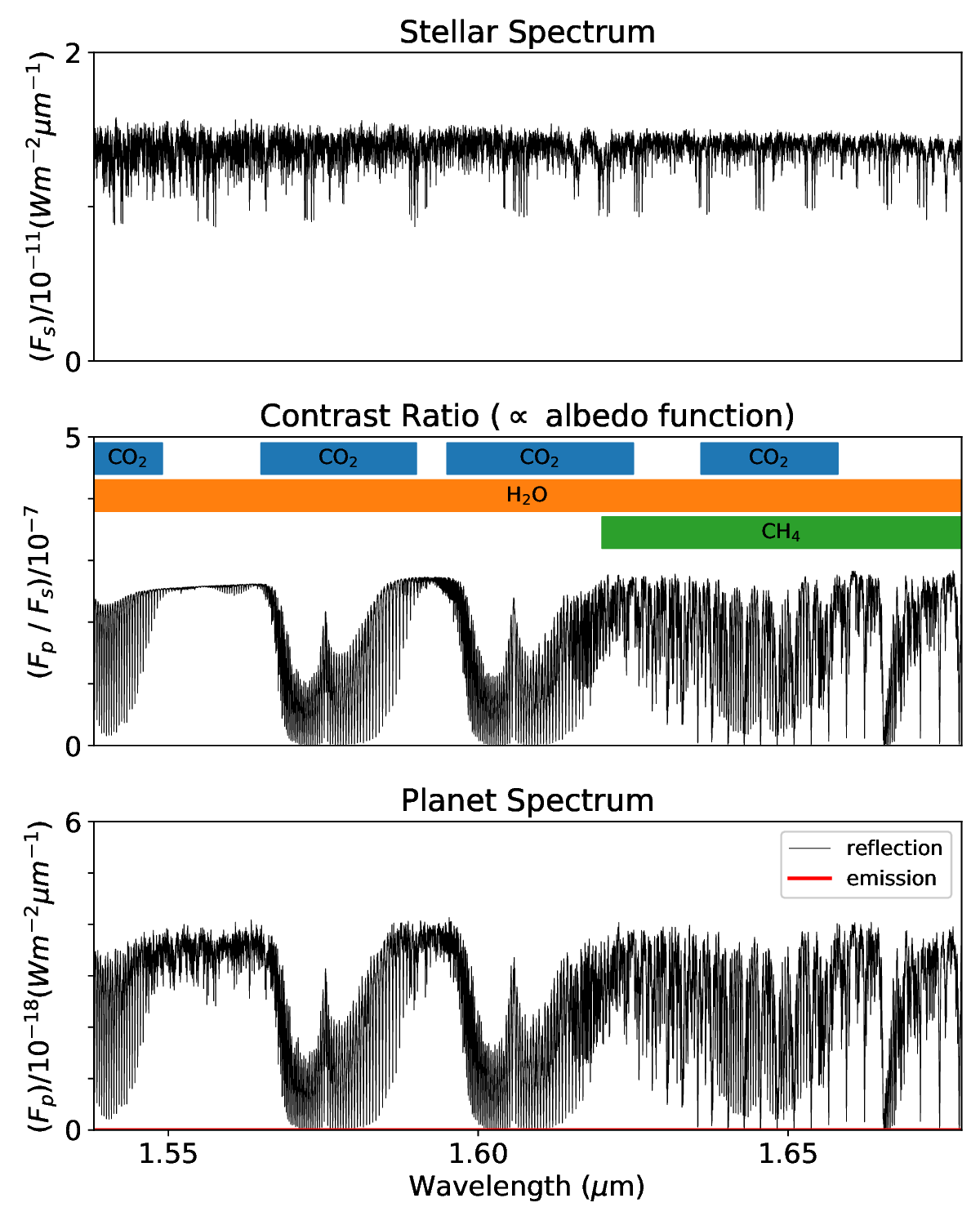}
    \caption{The spectra used in the simulation. \textbf{Top:} the stellar spectrum in units of intensity at Earth. \textbf{Middle:} the ratio of the exoplanet and stellar spectra ($4\times10^{-7}\sim17$mag). The features caused by different molecules in the planet's atmosphere have been highlighted. \textbf{Bottom:} the exoplanet's spectrum in units of intensity at Earth. In red is the thermal emission of the planet, which is virtually $0$ at approximately $2.7\times10^{-7}$ per cent of the planet's flux at these wavelengths assuming an equilibrium temperature of $\SI{234}{\kelvin}$. }
    \label{fig:spectra}
\end{figure}

Where $g(\alpha)$ is the phase function and $\alpha$ is the phase angle. We use a reflection spectrum which has been computed specifically for Proxima b from the Carl Sagan Institute \citep[]{Lin2020}. We use the Earth-like 1 bar oxic atmosphere model which assumes Earth-like mixing ratios for atmospheric gasses except for CO$_2$ which is more abundant by a factor of 100 to keep the planet from freezing. This model also assumes an Earth-like albedo for reflection of the planet's surface and accounts for the higher planet mass, lower instellation, and different stellar spectral type compared to the Earth. The average geometric albedo over wavelength ($1.538$ -- $1.678\SI{}{\micro\meter}$) of this model is $0.23$. \citeauthor{Lin2020} used a PHOENIX model \citep[]{Husser2013} for the host star Proxima Centauri when computing the spectrum of Proxima b so we also use a PHOENIX model for the star with the parameters in Table \ref{tab:proxima params}. The spectra of Proxima Centauri and Proxima b are shown in Fig. \ref{fig:spectra}. 

The simulations include the Doppler shift, $v_{Dop}$, resulting from the barycentric, systemic, and orbital velocities of each object. Additionally, we multiply the planet's flux by the illuminated fraction of the planet which is described by the phase function shown in Equation \ref{equ:ga}. This is simpler than Lambertain scattering which is typically assumed \citep[e.g.][]{Carron-Gonzalez2021, Spring2022} and results in the planet being slightly brighter at quadrature phases. A simpler function was chosen as the albedo and scattering properties of Proxima b are unknown at this time. Additionally, we do not include the rotational broadening, $\nu_{broad}$, of Proxima b's spectrum as, due to the planet's orbital period and slow stellar rotation period, the effect is insignificant \citep[]{Spring2022}.

%%%%%%%%%%%%%%%%%%%%%%%%%%%%%%%%%%%%%%%%%%%%%%%%%%
\subsection{Selecting Observation Times}
\label{sec:scheduler}

Our simulations model the orientation of the Proxima system so the dates of the simulated observations must be picked considering the observability of the planet just like real observations. To identify suitable dates, we assume we apriori know Proxima b's orbit (see Table \ref{tab:proxima params}) and check the following set of conditions at 10-minute intervals between $1^{\rm{st}}$ January 2030 and $1^{\rm{st}}$ January 2032:

\begin{enumerate}
    \item It must be nautical twilight or darker at Paranal.
    \item Proxima Centauri must be more than $\SI{45}{\degree}$ in elevation above the horizon at Paranal (airmass$<1.4$) for the AO to function well.
    \item Proxima b must not be behind the FPM at any time, as the throughput of the mask will reduce the reflected light signal too much for it to be detected.
    \item Proxima b must have at least half of its hemisphere illuminated as this leads to a higher reflected light signal. 
    \item Proxima b's velocity must be at least \SI{1}{\kilo\meter\per\second} different from Proxima Centauri's and Earth's as this prevents the spectral lines from being aligned which aids in the data reduction \citep[see e.g.][]{L2017}.
\end{enumerate}

We then identify a list of dates on which an observation of a given length can be made, if the length of the observation is less than the time window the planet can be observed, then the start time is chosen to minimise the airmass of the observation. Additionally, we also compute the total time Proxima b could be observed. Due to Proxima b's approximately $11$ day orbital period, it can remain widely separated from Proxima Centauri over a whole night. However, the FPM can severely limit the length of the observing window. First, the sky rotation changes the region of the sky covered by the mask which could result in the mask covering the planet part way through the night. Second, with a smaller mask the atmospheric dispersion can increase the amount of light around the mask edge which can impose stricter restrictions on the airmass in order to avoid persistence.

%%%%%%%%%%%%%%%%%%%%%%%%%%%%%%%%%%%%%%%%%%%%%%%%%%
\subsection{Modeling Earth's Atmosphere}
\label{sec:earthatm}

The Earth's atmosphere will contaminate the spectra we observe and modeling its effect is important in determining the feasibility of these observations. The airmass of the system is calculated using the known on sky coordinates of Proxima Centauri which is used to calculate the telluric transmission using TelFit \citep[]{telfit} and the telluric emission using SkyCalc \citep[]{skycalc12, skycalc13}. SkyCalc was not used for the telluric transmission as the discontinuities due to airmass interpolation caused artefacts in the simulation. The tellurics can be created for different weather conditions however, all the simulations presented here assume a surface pressure, surface temperature and humidity as given in Table \ref{tab:sim params} which are consistent with average conditions at Paranal and therefore very similar to those at Armazones, the ELT site.

%%%%%%%%%%%%%%%%%%%%%%%%%%%%%%%%%%%%%%%%%%%%%%%%%%
\subsection{Modeling HARMONI}
\label{sec:harmoni}

Observing Proxima b will push HARMONI to the limits, therefore a detailed instrument simulation is required to determine the feasibility of such observations. In this work we account for the throughput and emissivity of the ELT, the PSF of the combined ELT and HARMONI optics, the residual atmospheric dispersion, the FPM, the throughput and emissivity of HARMONI, and several noise sources from the detector and optics.

\subsubsection{The throughput and emissivity of the ELT optics} 
Our simulation uses the pre-calculated transmission as a function of wavelength from HSIM v310 while the thermal emission is assumed to be an non-ideal emitter i.e. a greybody with an emission temperature of $\SI{273}{\kelvin}$ and an emissivity equal to one minus the transmission of the ELT.

\subsubsection{The PSF of the combined ELT and HARMONI optics} 
We use the method presented in \citet{Fetick2018} to compute the long exposure PSF \citep[$>\SI{10}{\second}$ to average highly time-variable phase aberrations c.f.][]{Fetick2018, Fetick2019} for a given seeing and wavelength using the Power Spectral Density (PSD) computed in \citet{Houlle2021}. The sum of the PSF, sampled with HARMONI's spatial sampling, is normalised to unity. To reduce computational requirements, we assume a constant seeing of $\SI{0.57}{\arcsecond}$ (occurs approximately 30 per cent of the time at Armazones\footnote{\url{https://www.eso.org/sci/facilities/eelt/docs/ESO-193696_2_Observatory_Top_Level_Requirements.pdf}}) and only generate the PSF for the central wavelength, ignoring the wavelength dependence. As the planet resides close the mask edge, better seeing are desirable to minimise contamination of the exoplanets spectrum. We use the same realisation of the PSF for each observation. This results in a more optimistic and well behaved PSF than would likely be achieved on sky. \citet{Houlle2021} have a more realistic treatment of the PSF and so in Section \ref{sec:houlle} we compare our results to theirs.

\subsubsection{The residual atmospheric dispersion} 
HARMONI's HCAO mode uses an ADC with an optimal correction angle of $\SI{32.6}{\degree}$ which will only partially correct the atmospheric dispersion. Therefore, there will still be dispersion in the altitude of a point source's position which we calculate using the equations presented in \citet{Schubert2000}. 

\subsubsection{The focal plane mask}
Table \ref{tab:sim params} lists the shapes of the currently planned FPMs. To generate new FPMs to study in our simulation, we calculate the fraction of a mask of a given shape covering each spaxel. This fraction is then multiplied by the throughput ($10^{-4}$) to create the template for the mask. 

\subsubsection{The throughput and emissivity of HARMONI}
We compute the throughput and emissivity of each of the HARMONI components using information from HSIM v310. The total throughput is the product of the throughput of the all individual components. Additionally, using the throughput and emissivity at the component level, the total thermal emission seen by the detector is computed. 

\subsubsection{Noise sources}
The reflected light of Proxima b will be below the noise level in our simulation so we include a number of noise sources that might affect the recovery of the exoplanet's signal. The main source of noise comes from counting statistics i.e. \textit{Poisson noise}. To calculate this, we include the pre-calculated quantum efficiency values in HSIM v310 which vary slightly with wavelength as the detectors response is not perfectly uniform and assume the gain is unity. Additional noise sources included (see Table \ref{tab:sim params}) are i) \textit{the crosstalk}; caused by charge leakage from neighbouring pixels, ii) \textit{the read noise}; caused by noise in the electronics, iii) \textit{the dark current}; caused by small currents in the detector present even when it is not exposed to light, and iv) \textit{the thermal noise}, caused by the thermal emission of several components within view of the detector that are not along the optical path and so not included in the emission of HARMONI (computed as in HSIM v310 assuming greybodies for each of the components in view). In our simulations where integration time is limited by the persistance limit, the Poisson noise and read noise contribute similar amounts to the noise on the exoplanet's spectrum.

\subsubsection{Effects not included}
In our simulations, we neglect the non-linear relationship between the number of photons absorbed by a pixel and the charge on that pixel as the saturation limit is reached. Proxima b's spectrum is not located on or near a pixel reaching this limit however, if significant saturation occurs, Proxima b's spectrum will be affected although this is not modelled in our simulations. We also neglect the persistence of the detector which is the limit beyond which the charge on a pixel cannot be completely discharged during a detector read. This causes a residual charge in the pixel which can last for the rest of the observation and possibly affect subsequent observations. However, all the simulations presented here are kept below the limit where persistence is noticeable. Lastly, the generated PSF does not include the additional scattering caused by the sharp edge of the focal plane mask. HARMONI's optics past the focal plane mask are significantly oversized, approximately 15 times larger than necessary, at the spatial scale used in this work, in order to accommodate the settings with larger spaxel scales. Thus, although the finite size of the optics means that, in principle, some of the scattered light is not reimaged back to the mask edge, the additional background contamination is not expected to be significant for HARMONI’s high contrast mode. A quantitative estimate of the magnitude of contamination is complex and is beyond the scope of this work.

%%%%%%%%%%%%%%%%%%%%%%%%%%%%%%%%%%%%%%%%%%%%%%%%%%

\begin{table}
\begin{center}   
    \begin{tabular}{m{0.20\textwidth}m{0.18\textwidth}m{0.02\textwidth}}
        \hline
        Instrumental Mode & Value & Ref \\
        \hline
        Spaxel Scale & $\SI{3.88}{\mas}$ & $1,2$ \\
        Spectral Resolving Power & $17385$ & $1,2$ \\
        Wavelength range & $1.538$ -- $1.678\SI{}{\micro\meter}$ (H-High) & $1,2$ \\
        AO mode & HCAO - SP1 apodizer & $3$ \\
        \hline
        Observing Conditions & Value & Ref \\
        \hline
        Longitude & $-70^{\circ}24'18''$ & $4$\\
        Latitude & $-24^{\circ}37'39''$ & $4$\\
        Airmass & $1.25$--$1.4$ & -\\
        Seeing & $\SI{0.57}{\arcsecond}$ (fixed) & $4,5$\\
        Surface Pressure & $\SI{795}{\hecto\pascal}$ & $4$ \\
        Temperature & $\SI{283}{\kelvin}$ & $4$ \\
        Relative Humidity & $20$ per cent & $4$ \\
        \hline
        Instrumental Parameters & Value & Ref\\
        \hline
        ELT Transmission$_{\rm{Havg}}$ & $0.755$ & $2$ \\
        ELT Emission$_{\rm{Havg}}$ & $\SI{3.8}{\photon\per\second\per\square\meter\per\micro\meter\per\square\arcsec}$ & $2$ \\
        Atmospheric Dispersion \newline Corrector angle & $\SI{32.6}{\degree}$ (fixed) & $3$ \\
        Focal Plane Mask transmission & $10^{-4}$ & $3$ \\
        HARMONI Transmission$_{\rm{Havg}}$ & $0.438$ & $2$ \\
        HARMONI Emission$_{\rm{Havg}}$$\dagger$ & $\SI{0.5}{\photon\per\second\per\square\meter\per\micro\meter\per\square\arcsec}$ & $2$ \\
        Total Throughput$_{\rm{Havg}}$$\ast$ & $0.30$ & $-$ \\
        Quantum Efficiency$_{\rm{Havg}}$ & $0.9$ & $2$ \\
        Crosstalk & $0.02$ per adjacent pixel & $2,3$ \\
        Read Noise & $12 e^-$ per pixel & $2,3$ \\
        Dark Current & $0.0053 e^- s^{-1}$ per pixel & $2,3$ \\
        Thermal Noise from Cryostat & $0.017 e^- s^{-1}$ per pixel & $2$ \\
        Persistence Limit & $30000 e^-$ & $2$ \\
        \hline
        Focal Plane Masks & Value & Ref \\
        \hline
        FPM for SP1 apodizer (H band) & ellipse: $\SI{50}{\mas}$ x $\SI{58}{\mas}$ & - \\
        FPM for SP1 apodizer (K band) & ellipse: $\SI{72}{\mas}$ x $\SI{76}{\mas}$ & - \\
        FPM for SP2 apodizer & ellipse: $\SI{96}{\mas}$ x $\SI{96}{\mas}$ & - \\
        \hline
    \end{tabular}
\caption{Parameters used to simulate observations with HARMONI HCAO mode and accompanying references. See Table \ref{tab:proxima params} for the parameters used to simulate the Proxima Centauri system. \newline
        Havg indicates wavelength dependant quantities that have been averaged over the H-High band in this Table. The wavelength dependence is included in the simulation.
        $\dagger$ This assumes the focal plane relay has a temperature of $\SI{-10}{\celsius}$. In the current design, this temperature has increased to $\SI{+2}{\celsius}$ however this difference will not make an appreciable change to the noise and therefore the results presented here. 
        $\ast$ This excludes the focal plane mask.\newline
        $1$ \citet{HARMONI}
        $2$ using information from HSIM v310 \citet{Zieleniewski2015}
        $3$ \citet{Houlle2021}
        $4$ \url{https://www.eso.org/sci/facilities/paranal/astroclimate/site.html}
        $5$ \url{https://www.eso.org/sci/facilities/eelt/docs/ESO-193696_2_Observatory_Top_Level_Requirements.pdf}
        }
\label{tab:sim params}
\end{center}
\end{table}

%%%%%%%%%%%%%%%%%%%%%%%%%%%%%%%%%%%%%%%%%%%%%%%%%%
\subsection{Simulating the Observations}
\label{sec:subsims}

Ideally, each detector integration would be simulated separately as in \citet{Houlle2021} however this is very computationally intensive. In this work, we simulate groups of detector integrations creating one output observation for each group e.g. a group of $60$ detector integrations, which are each $\SI{60}{\second}$, becomes a single $1$ hour observation. By simulating groups, hereafter referred to as `sub-simulations', we reduce the computation time and final data volume compared with simulating each detector integration separately. The noise is scaled appropriately such that it is equivalent to the noise (including read noise) that would be present if each detector integration was simulated separately. Due to computational limitations, the same realisation of the PSF with $\SI{0.57}{\arcsecond}$ seeing is used for each observation. The sub-simulations account for time dependent changes in the motion of the planetary system and the airmass dependence in the tellurics making them a reasonable approximation to simulating all the detector integrations individually.
It should be noted that no de-rotation is performed during the sub-simulation so we limit the length of our sub-simulations to $1$ hour ($30$ minutes in Section \ref{sec:molecule mapping}) to prevent the exoplanet's signal from being smeared out by more than $3$ spaxels. These observations assume perfect flat fielding and wavelength calibration.

%%%%%%%%%%%%%%%%%%%%%%%%%%%%%%%%%%%%%%%%%%%%%%%%%%
%%%%%%%%%%%%%%%%%%%%%%%%%%%%%%%%%%%%%%%%%%%%%%%%%%
%%%%%%%%%%%%%%%%%%%%%%%%%%%%%%%%%%%%%%%%%%%%%%%%%%

%%%%%%%%%%%%%%%%%%%%%%%%%%%%%%%%%%%%%%%%%%%%%%%%%%
\section{Using the Molecule Mapping Technique}
\label{sec:molecule mapping}
To first demonstrate the efficacy of our simulation, we simulate an optimally located exoplanet, assumed to have the same parameters as those in Table \ref{tab:proxima params} but with a semi-major axis two times that of the real Proxima b, $\SI{0.097}{\astronomicalunit}$ ($\SI{74.6}{\mas}$), to put the exoplanet in the centre of the dark annulus (see Fig. \ref{fig:example_sim}). This is not a physically realistic system since the exoplanet's model spectrum is calculated for a planet with a semi-major axis of $\SI{0.0485}{\astronomicalunit}$ ($\SI{37.3}{\mas}$) but it serves as a demonstration of the technique.

%%%%%%%%%%%%%%%%%%%%%%%%%%%%%%%%%%%%%%%%%%%%%%%%%%
%%%%%%%%%%%%%%%%%%%%%%%%%%%%%%%%%%%%%%%%%%%%%%%%%%
\subsection{Fiducial Planet}
\label{sec:example}
\begin{figure}
	\includegraphics[width=\columnwidth]{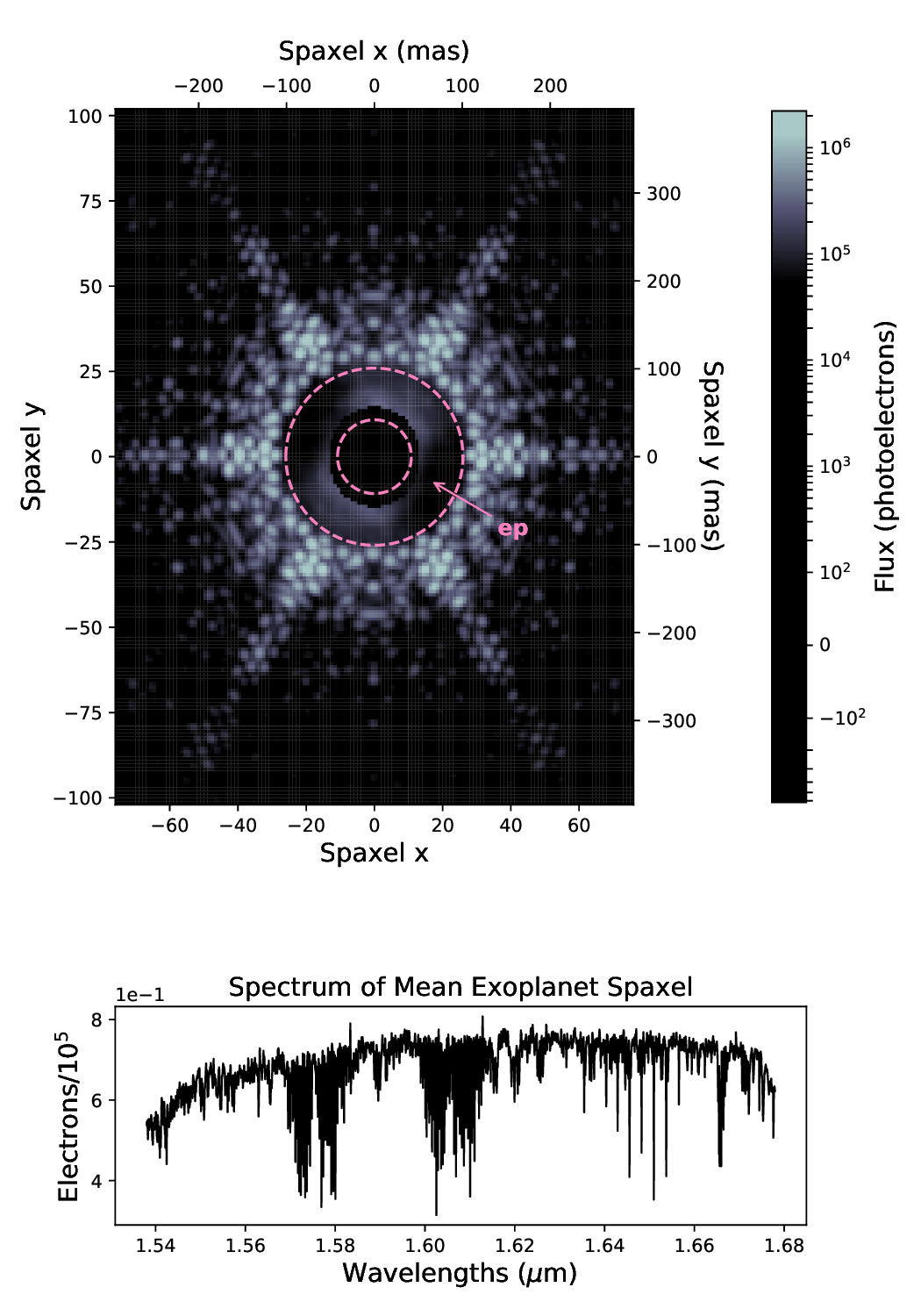}
    \caption{Fiducial simulation made with the currently planned FPM (the black ellipse in the centre of the top panel). The top panel shows the on-sky image at a single wavelength for $3$ hours of observations. The simulation contains an exoplanet at a separation of $\SI{68.3}{\mas}$ -- it is not at quadrature so its separation is less that $\SI{74.6}{\mas}$ -- from the star which puts it between the inner and outer working angles of the apodizer (dashed lines). We note the FPM is slightly larger than the inner working angle of the apodizer. The mean position of the planet is indicated be the `ep' label but the planet is too dim to be seen directly and is spread out due to sky rotation. The white diffuse emission inside the dark annulus is the wind-driven halo. The lower panel shows the spectrum of the `ep' spaxel. It is largely dominated by tellurics and the stellar spectrum.}
    \label{fig:example_sim}
\end{figure}

We simulate observations with a spaxel size of $\SI{3.88}{\mas}$, using the H-High band grating of HARMONI at the highest spectral resolving power $(R = 17385)$, and with the smallest FPM for SP1 apodizer (H band) which is $\SI{50}{\mas}$ x $\SI{58}{\mas}$ in size. The detector integration time is $\SI{60}{\second}$ so that the image is near but not at the persistence limit. A total integration time of $\SI{3}{\hour}$ is simulated per night, excluding overheads, which requires $6$ sub-simulations of $30$ detector integrations ($\SI{0.5}{\hour}$ each). Five nights of data are simulated, see Table \ref{tab:fid params} for details. It is worth noting that higher illumination fractions can only be observed at smaller star-planet separations resulting in a trade off between increased flux and increased contamination. The simulation for the 2$^{\rm{nd}}$ of April, with all sub-simulations stacked, is shown in Fig. \ref{fig:example_sim}.

\begin{table}
\begin{center}   
    \begin{tabular}{m{0.08\textwidth}m{0.13\textwidth}m{0.2\textwidth}}
        \hline
        Date & Average Separation & Average Fractional Illumination \\
        \hline
        2030-04-02 & $\SI{68.3}{\mas}$ & $0.70$ \\
        2030-04-03 & $\SI{74.5}{\mas}$ & $0.52$ \\
        2030-04-09 & $\SI{74.0}{\mas}$ & $0.56$ \\
        2030-04-10 & $\SI{65.5}{\mas}$ & $0.74$ \\
        2030-04-13 & $\SI{65.9}{\mas}$ & $0.73$ \\
        \hline
    \end{tabular}
\caption{The separations and fractional illuminations of the fiducial planet during the simulated observations.
        }
\label{tab:fid params}
\end{center}
\end{table}

%%%%%%%%%%%%%%%%%%%%%%%%%%%%%%%%%%%%%%%%%%%%%%%%%%
\subsection{Removing Contaminating Spectra}
\label{sec:datareduction}

\begin{figure}
    \centering
	\includegraphics[width=\columnwidth]{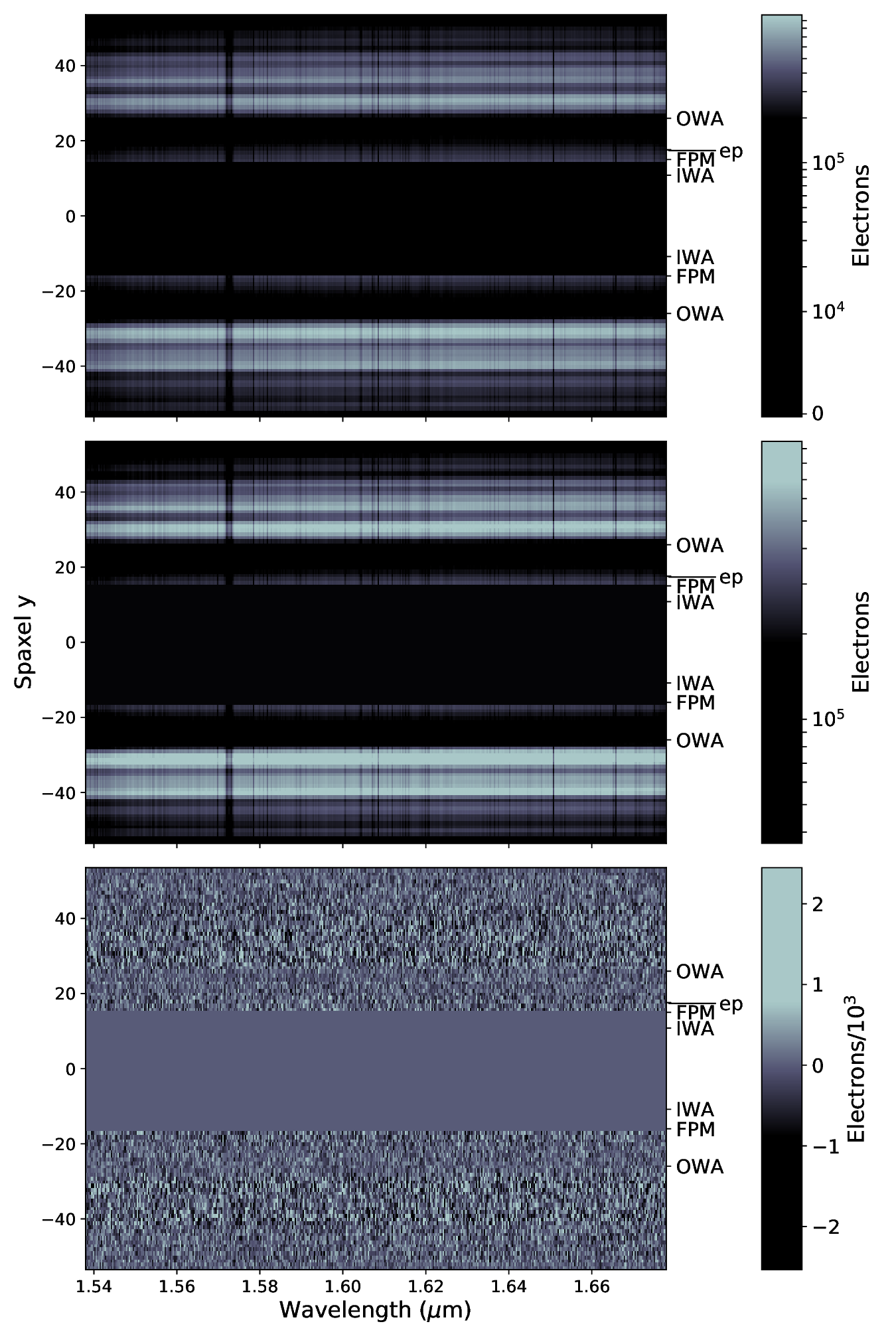}
    \caption{Data reduction applied to the fiducial simulation. 
    The plot shows, at each stage in the data reduction, stacked sub-simulations residuals for one night ($3$ hours of integration time) for a slice along the spaxel $x=0$. These have been de-rotated for visual purposes only so the exoplanet's location is the same throughout the observation thereby showing the residuals at the exoplanet's location.  The top panel shows the original flux; the middle shows the residual after the cross talk correction and background subtraction, and the bottom panel shows the residuals after the data reduction from \citet{HH2018-2} is performed. The bottom panel appears to only contain residual noise and the exoplanet's spectrum is not visible. Along the side of each panel, the locations of the inner and outer working angles of the apodizer (IWA and OWA respectively), the edges of the FPM and the position of the planet (ep) are indicated. The increase in the standard deviation in the noise outside the OWA is due to the increase in speckles. The region covered by the FPM is masked in the middle and bottom panels. Note the change in residual colour bar scales.}
    \label{fig:example_reduction}
\end{figure}

Despite the apodizer's reduction of the stellar PSF, the exoplanet remains too faint to be directly visible. The stellar and telluric contamination were removed from each sub-simulation using the following steps. First, the effect of crosstalk was removed to first order by subtracting the original spectrum of each adjacent pixel on the detector multiplied by the crosstalk ($0.02$). Without this first step, a detection is not possible due to significant residual stellar contamination. Next, the background spectrum was removed by subtracting the median spectrum of the $1,000$ spaxels with the lowest total flux (excluding spaxels within the FPM). Finally, the stellar and telluric contamination was removed using the method in \citet{HH2018-2} which subtracts the smoothed continuum of each spaxel multiplied by mean spectrum of the $10,000$ spaxels with the highest flux. Such a large number was chosen through trial an error as it gave the best signal-to-noise out of the parameters tested. The data reduction as applied to the fiducial simulation is shown in Fig. \ref{fig:example_reduction}.

%%%%%%%%%%%%%%%%%%%%%%%%%%%%%%%%%%%%%%%%%%%%%%%%%%
\subsection{Cross Correlation Analysis}
\label{sec:crosscorrelation}

To recover the signal of the exoplanet in these reduced data we use cross-correlation analysis which combines the exoplanet's signal spread out across wavelength. The cross-correlation coefficient is a measure of the degree of correlation (similarity) of the Doppler-shifted model and the reduced spectrum, like a normalised integration of the signal of the exoplanet over wavelength. Therefore, a spectrum containing only noise should produce a smaller correlation coefficient than one containing the exoplanet's spectrum. For reflected light spectra, the exoplanet's spectrum contains the stellar spectrum so the cross-correlation method is also sensitive to any residual stellar spectrum in these reduced data. 

Each spaxel in the reduced simulation is cross-correlated with the model of the exoplanet's spectrum described in Section \ref{sec:spectra}. We use the Pearson cross-correlation coefficient which is defined as:
%\ud{\citep[modified from][]{RF2013}}{}:

\begin{equation}
    C(v) = \frac{\sum_{\lambda} (f_\lambda(v) - \bar{f}(v)) (s_\lambda - \bar{s})}{\sqrt{ \sum_{\lambda} (f_\lambda(v) - \bar{f}(v))^2  \sum_{\lambda} (s_\lambda - \bar{s})^2 }}.
\end{equation}

It is a function of the Doppler shift, $v$, of the model spectrum $f$. Here, $s$ is the reduced spectrum of a spaxel in the sub-simulation. The sum $\lambda$ is over the wavelength bins and $\bar{f}(v)$ and $\bar{s}$ indicate the average over wavelength of the model spectrum and reduced spectrum respectively.

\begin{figure}
	\includegraphics[width=\columnwidth]{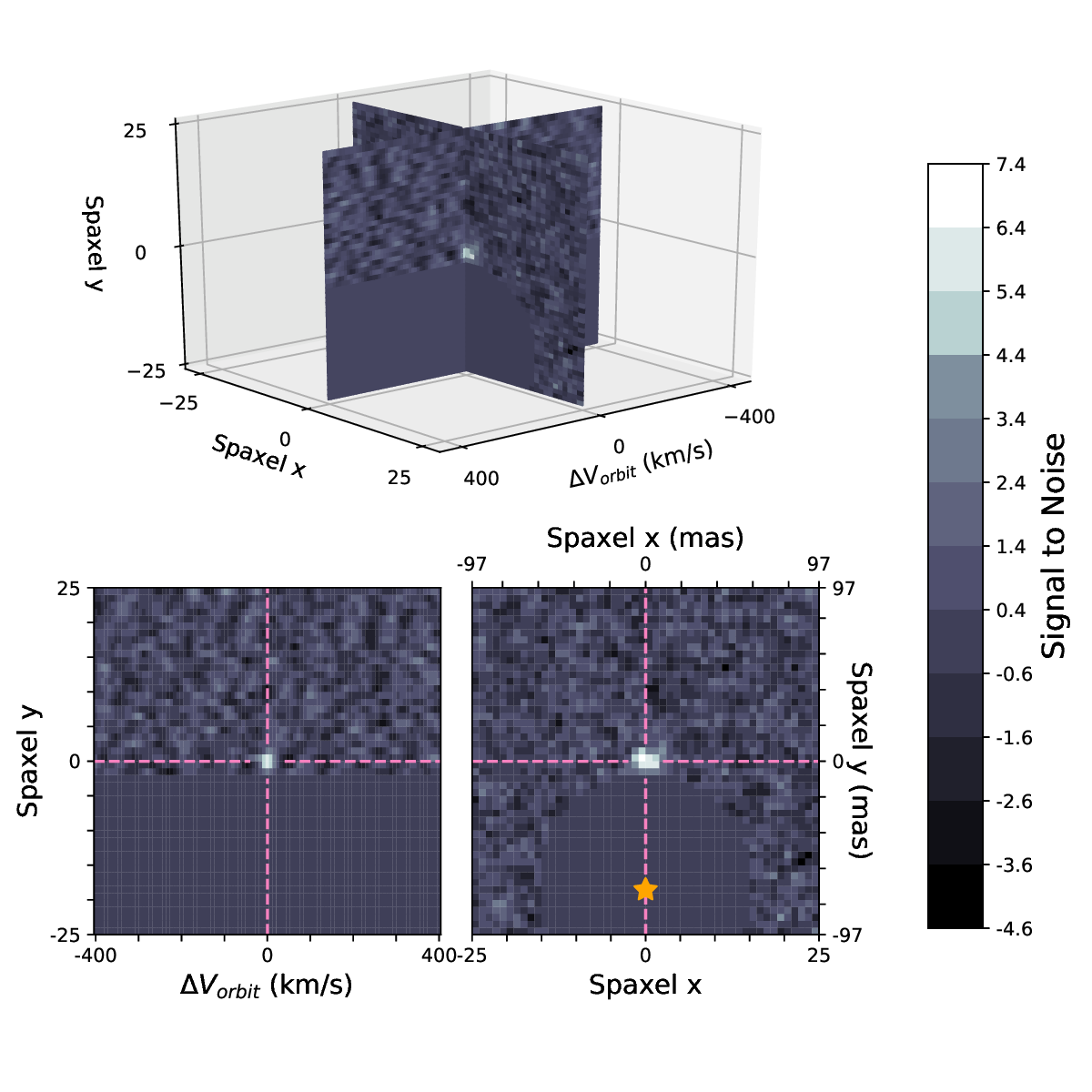}
    \caption{Signal-to-noise of the fiducial planet simulation. The top panel shows two slices of the CCF cube, one in the plane on the sky and one along the velocity axis. The bottom panels show the same two slices separated for clarity with pink dashed lines indicating the expected position of the signal. The region covered by the FPM is greyed-out. The bottom left panel shows the signal-to-noise for the spaxels along spaxel $x=0$ for different velocities relative to the exoplanet's velocity. The bottom right panel shows the signal-to-noise of all the spaxels at a Doppler shift equal to the exoplanet's velocity. The yellow star in the bottom right panel indicates the location of the star. Note that this injected fiducial exoplanet does not represent Proxima b. }
    \label{fig:example_cc}
\end{figure}

We cross-correlate each of the sub-simulations at an array of Doppler shifts centered on the injected planet using the same model as the cross-correlation template yielding a 3D data cube of cross-correlation coefficients (hereafter `CCF cube'). These are then de-rotated so the exoplanet is in the same location in the interpolated grid and then added together to create the final CCF cube for the full integration time. This analysis requires knowledge of the planet's orbit, however, as discussed in Section \ref{sec:unknownorbit}, if the orbit is unknown but all the observations are taken at approximately the same point in the planet's orbit then the analysis can proceed as described here. This cube is converted to signal-to-noise by dividing the cross-correlation coefficients of each spaxel by the standard deviation of the coefficients. The standard deviation is calculated for each spaxel separately excluding velocities expected to include the main peaks in the model's auto-correlation function. For these observations, this is a continuous region in velocity space $\SI{168}{\kilo\meter\per\second}$ wide centered on the velocity of the planet. Shown in Fig. \ref{fig:example_cc} are two slices of the final CCF cube created from the fiducial simulation. A signal with $S/N = 7.4$ is seen at the exoplanet's expected position and velocity, confirming we can recover planets of this contrast in the simulation. The signal is slightly spread out in the $x$ axis due to the sky rotation in the sub-simulations.

%%%%%%%%%%%%%%%%%%%%%%%%%%%%%%%%%%%%%%%%%%%%%%%%%%
\subsection{Comparison with Previous Work}
\label{sec:houlle}
\begin{figure}
	\includegraphics[width=\columnwidth]{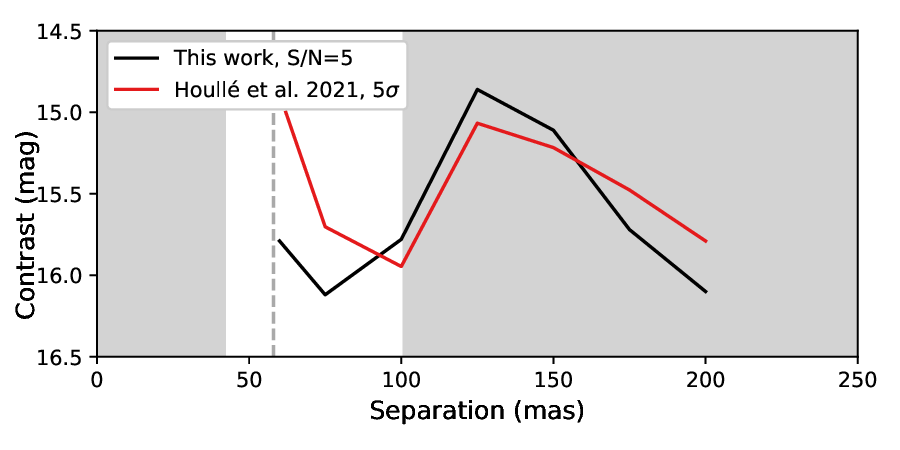}
    \caption{A contrast curve for an $S/N = 5$ detection of a T-type companion (T$_{\rm{eff}}=\SI{800}{\kelvin}$) with $2$ hours of integration time (black) using the simulation process and analysis described in this work and its comparison $5 \sigma$ detection (red) curve from \citet{Houlle2021}. The x axis indicates the separation from the star in the centre of the field of view. The shaded regions are outside the inner and outer working angles for the apodizer. The focal plane mask is $\SI{58}{\mas}$ at its widest point, indicated by the dashed line, so we do not attempt to recover planet's closer than this. The curves cannot be directly compared as they use different signal-to-noise metrics although the scales and shapes broadly agree indicating our simulation is a reasonable approximation of previous work. The reduced detection efficiency around $\SI{125}{\mas}$ is due to increased speckle noise at these separations which can be seen in Fig. \ref{fig:example_sim}.}
    \label{fig:hcomp}
\end{figure}

As mentioned in Section \ref{sec:harmoni}, our simulation uses an idealised PSF and simulates groups of integrations (sub-simulations) rather than each detector integration separately. These simplifications are not made in \citet{Houlle2021} so a comparison between the two should in principle indicate whether they will significantly affect our results. Unfortunately, a direct comparison is not possible as we are unable to use their signal-to-noise metric as it requires individual detector integrations to be simulated. In addition, the signal-to-noise recovered could be affected by differences in the data reduction as, for example, we do not use principal component analysis. We create simulations using the same spectral models \citep[ATMO model at T$_{\rm{eff}}=\SI{800}{\kelvin}$, log(g)$=4.0$;][]{Philips2020}, Doppler shifts and airmasses as in \citet{Houlle2021}, and compute a $S/N = 5$ detection contrast curve for $2$ hours of integration time using our metric. Our curve and its equivalent from \citet{Houlle2021} for a $5 \sigma$ detection with $2$ hours of integration time are shown in Fig. \ref{fig:hcomp}. Due to the different metrics being uses, the absolute values should not be compared, however the similarity in the shape and scale indicates that our simulation is valid and a reasonable approximation of previous work. We note, however, these observations will be pushing the limits of what HARMONI could achieve and it is ultimately very hard to predict how an instrument will behave prior to its operation. Ideally observations of wider-separated and brighter planets would be used to determine HARMONI's performance prior to attempting these observations, however, any changes in instrument design required to facilitate these observations would have to be committed to hardware before HARMONI's true performance is known.

%%%%%%%%%%%%%%%%%%%%%%%%%%%%%%%%%%%%%%%%%%%%%%%%%%
%%%%%%%%%%%%%%%%%%%%%%%%%%%%%%%%%%%%%%%%%%%%%%%%%%
%%%%%%%%%%%%%%%%%%%%%%%%%%%%%%%%%%%%%%%%%%%%%%%%%%
\section{Observing Proxima b with HARMONI}
\label{sec:results}
With molecule mapping, it is possible to push to within the inner working angle \citep[e.g.][]{HH2018-2} ($\sim\SI{40}{\mas}$ or $5 \lambda/D$ at $\SI{1.538}{\micro\meter}$ for the SP1 apodizer), however even the smallest of the FPMs ($\SI{50}{\mas}$ by $\SI{58}{\mas}$) will completely cover the orbit of Proxima b which has a maximum separation from Proxima Centauri of $\SI{37.3}{\mas}$. It is not possible to detect Proxima b when it is behind the FPM and, due to limited space, it is not possible to add additional apodizers and FPMs to HARMONI. We suggest and simulate here two solutions to this problem: offsetting the FPM, or replacing one of the FPMs with a smaller mask. For the latter case, a careful assessment on the impact of this change to other science cases compared with the benefits to this case would be needed to justify such a change.

It is possible that Proxima b's orbit will be well constrained before HARMONI observes it, therefore, in our simulations, we assume we already know the orbit of Proxima b (see Table \ref{tab:proxima params}) meaning we can predict it's on-sky position and Doppler shift. Proxima b is observable in all the observations simulated and the final CCF cubes are Doppler shifted and de-rotated to align the planet's signal. Later in Sections \ref{sec:timeavalible} and \ref{sec:unknownorbit} we explore when such observability would occur. 

%%%%%%%%%%%%%%%%%%%%%%%%%%%%%%%%%%%%%%%%%%%%%%%%%%
\subsection{Offsetting the Mask}
\label{sec:offset}
To observe Proxima b with the current instrument design, the star could be offset from the centre of the field of view so that the mask does not cover the on sky location of Proxima b. This should not affect the performance of the AO system. We create a simulation where the star is offset by $\SI{20}{\mas}$ in azimuth (see panel 4 of Fig. \ref{fig:leakage}). We assume we know the orbit of Proxima b and simulate $70$ observations -- each with a total of one hour integration time using a detector integration time of $\SI{60}{\second}$ -- spread out over two years in which Proxima b is not behind the focal plane mask.  We treat each observation separately in the data reduction due to the changes in relative positions and velocities of the star, planet and Earth. The analysis is performed as in Section \ref{sec:example} which produces $70$ CCF cubes. We de-rotate and Doppler shift these, assuming the known orbit, to align the exoplanet's signal.

%%%%%%%%%%%%%%%%%%%%%%%%%%%%%%%%%%%%%%%%%%%%%%%%%%
\subsection{Decreasing the Mask Size}
\label{sec:masksize}
Ideally the orbit of Proxima b would need to be known so that the direction and magnitude of the offset could be calculated. If the orbit is unknown a guess for the offset could be used but this increases the number of observations required for a detection as discussed later in Section \ref{sec:selectmask}. An alternative would be to replace one of the FPMs with a smaller mask that does not completely cover Proxima b's orbit. This could be accomplished during construction or operation; however, the latter would likely only to occur during scheduled interventions approximately 5-10 years after first light. 

\begin{figure*}
	\includegraphics[width=\textwidth]{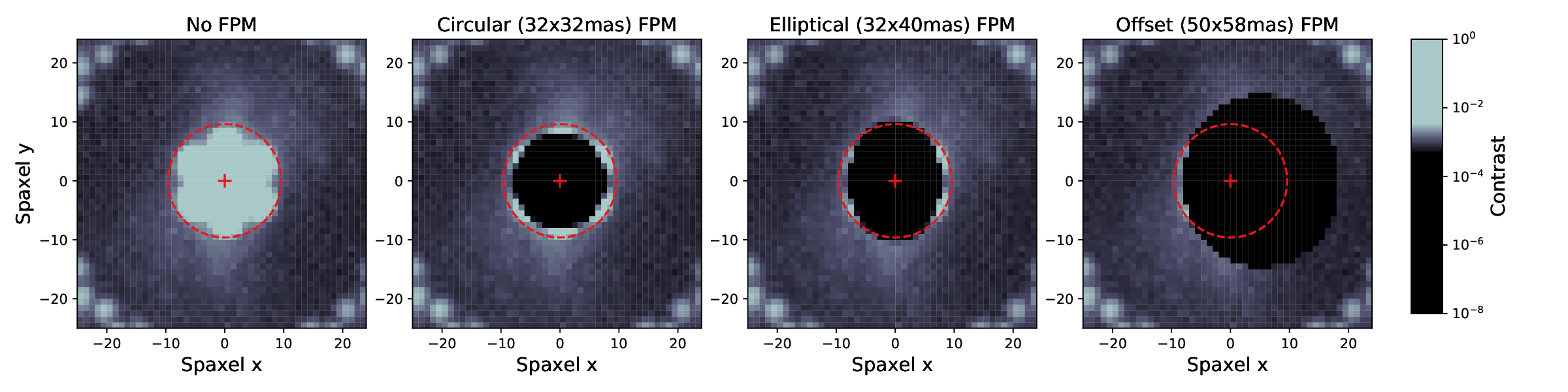}
    \caption{FPM's analysed in this work displayed on a map of the stellar PSF. The red cross indicates the position of the star and the red dashed line is the maximum separation of Proxima b, that is Proxima b will always lie somewhere within this circle. Starlight leaking around the edge of the mask will scatter and increase the stray light in the instrument. This is not modelled in our simulation.}
    \label{fig:leakage}
\end{figure*}

We analyse three different sizes and shapes for a new FPM to determine which would be best for observing Proxima b. We choose `No FPM' mask, a `Circular FPM with radius $\SI{32}{\mas}$' and an `Elliptical FPM with dimensions $\SI{32}{}\times\SI{40}{\mas}$'. The region covered by circular FPM will not change significantly which is ideal if the on-sky location of Proxima b is unknown but the elliptical mask better covers the central diffraction peak reducing the amount of star light leakage (see Fig. \ref{fig:leakage}). 

We create simulated observations for each of the FPMs assuming a detector integration time of $\SI{60}{\second}$ so that the detector is near but not at the persistence limit in the circular and elliptical cases. This restricts the airmass of the observations due to leakage around the circular mask caused by atmospheric dispersion. This could be improved by decreasing the integration time but doing so would make the read noise the dominant noise source. In the no FPM simulation, the core of the PSF is over the persistence limit. In total, $56$ hours of integration time are simulated, on the same dates in each case, spread out over two years. The times selected are primarily driven by the elliptical mask due to its larger size. The restrictions imposed by our observing conditions limit the continuous time window for observations to around $2$ hours. We analyse each sub-simulation as in Section \ref{sec:molecule mapping} which yields $56$ CCF cubes per FPM which we de-rotate and Doppler shift, assuming the known orbit, to align the exoplanet's signal.

%%%%%%%%%%%%%%%%%%%%%%%%%%%%%%%%%%%%%%%%%%%%%%%%%%
\subsection{Detecting Proxima b}
\label{sec:timeneeded}

\begin{figure}
	\includegraphics[width=\linewidth]{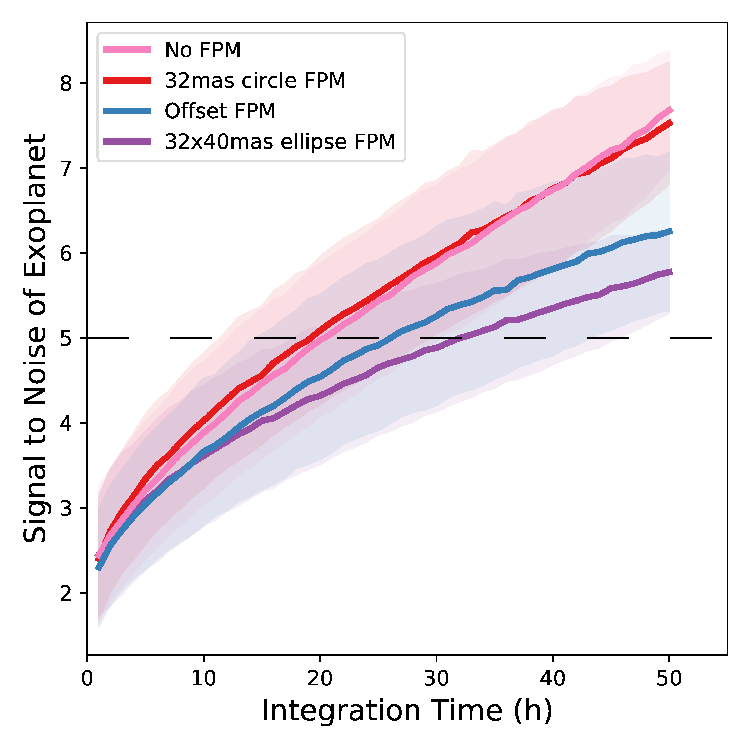}
    \caption{Average recovered signal-to-noise of Proxima b in our simulations using the planet absorption and stellar spectrum cross correlation model for each FPM as a function of integration time. A signal-to-noise of $5$ is indicated by the dashed line. The standard deviation in the recovered signal-to-noise is indicated by the shaded region. See Section \ref{sec:selectmask} for a discussion of the implications of using these options and ultimately which is the most suitable.}
    \label{fig:result}
\end{figure}

To calculate the integration time necessary for a detection with a given mask, the CCF cubes are stacked and the signal-to-noise of the stack is computed as in Section \ref{sec:crosscorrelation}. The signal-to-noise of the detection is taken as the maximum in the three-by-three grid of spaxels centred on the expected position of the exoplanet. The signal-to-noise for integration times less than the total time simulated is obtained by varying which cubes are stacked. This is repeated and the mean and standard deviation in the signal-to-noise recovered is shown for each of the FPM's in Fig. \ref{fig:result}. A signal-to-noise of $5$ is used as the detection threshold as, when no exoplanet signal is present, $S/N\approx4$ can still be obtained through the random combination of noise \citep[see e.g.][]{Cabot2019, Spring2022}. A detection with a $S/N\geq5$ will require at least $20$ hours of time on the ELT (for the assumed orbital orientation of Proxima b). Our simulations show significant variations in the signal obtained for different observations which leads to changes in the amount of time required. In the simulations, as little as $12$ or as much as $30$ hours could be necessary for a detection with the circular or no FPM, between $18$ to $45$ hours for the elliptical mask and between $14$ to $43$ hours for the offset mask. The variation in the signal to noise recovered is due to three effects. First, the exoplanet is at different on sky positions which changes the amount of stellar contamination. Second, the amount of Proxima b which is illuminated changes meaning the amount of light we receive from the planet is changing. Finally, the random nature of the noise can affect the signal-to-noise recovered even if everything else is kept the same. We note that the due to the difficulty with selecting common dates, the offset simulation is comprised of a different set of observations to the other three cases. However, the average separation -- $\SI{35.5}{\mas}$ for the offset mask and $\SI{36.9}{\mas}$ for the other three -- and illumination fraction -- $0.63$ for the offset mask and $0.56$ for the other three -- of Proxima b is similar in both cases. 

We note that the signal-to-noise of the detection does not scale with the square root of the integration time. This is due to a number of effects, the most significant of which are i) the read noise at the planet's location is similar in magnitude to the photon noise and, ii) there is residual stellar spectrum present in the noise which correlates with the planet's spectrum. Less significant effects include iii) the signal is smeared out slightly within a sub-simulation due to the on sky rotation, iv) the FPM's effect on the background spectrum which reduces the efficiency of the data reduction near the edge of the mask and, v) residuals from the telluric spectrum which can correlate with the planet's spectrum as they contain the same species. A better data reduction may improve the signal-to-noise recovered, however, perfecting the data reduction is beyond the scope of this work. Even in this case where we perfectly know and control all sources of noise, the simulation does not scale as Poisson statistics. Careful and detailed simulations, particularly when close to the read noise, are needed for all future instrumentation for the Extremely Large Telescopes to fully understand how they will respond for high resolution spectroscopy of exoplanet atmospheres. 

%%%%%%%%%%%%%%%%%%%%%%%%%%%%%%%%%%%%%%%%%%%%%%%%%%
\subsection{Detecting Proxima b's Atmosphere}
\label{sec:atmdect}

\begin{figure}
	\includegraphics[width=\linewidth]{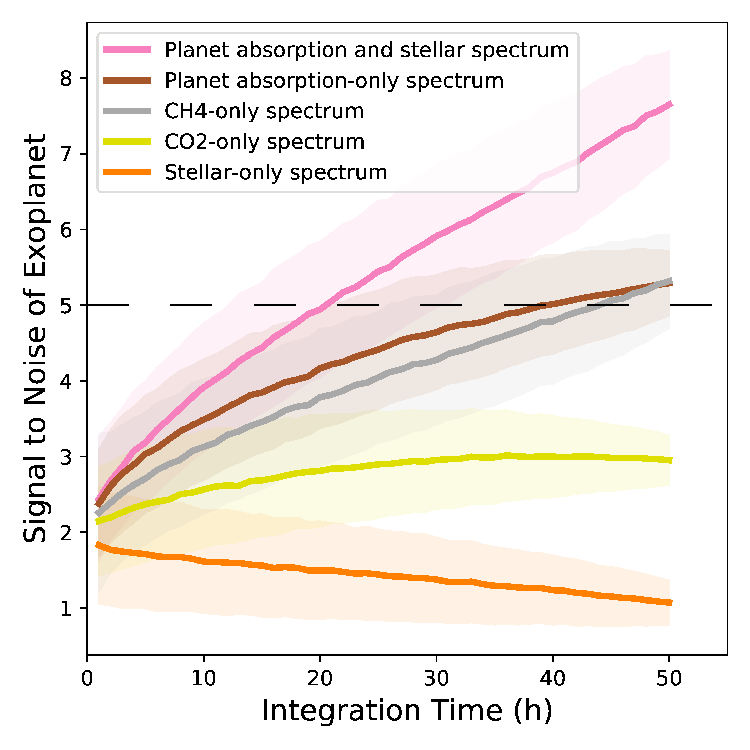}
    \caption{The average recovered signal-to-noise of Proxima b for the no FPM simulation as a function of integration time when cross-correlated with the stellar spectrum only model (orange), planet absorption only model (brown), the model planet spectrum (atmospheric absorption and reflected stellar lines, pink), a model containing only the CH$_4$ lines (grey) and a model containing only the CO$_2$ lines (yellow). The $1\sigma$ variation in the recovered signal-to-noise is indicated by the shaded region.}
    \label{fig:co2dect}
\end{figure}

In Section \ref{sec:timeneeded} we used a cross-correlation model that is a perfect match to the planet's spectrum in the simulated data, which contains features from the reflected stellar spectrum and from the atmospheric absorption of the planet. To determine if we are sensitive to the planet's atmospheric absorption, we repeat the `no FPM' simulation but cross correlate with a model of the stellar spectrum only (no planet absorption features) \citep[c.f.][]{Hawker2019} and with a model with only the planet absorption (no stellar features). We also cross correlate with a model containing only the CO$_2$ lines and one with only the CH$_4$ lines. A comparison between the signal-to-noise recovered for the different cross correlation models is shown in Fig. \ref{fig:co2dect}. When the stellar and the planetary absorption features are used as the cross correlation model, a detection with a signal-to-noise of $5$ is obtainable in approximately $20$ hours which increases to $40$ hours if only the planet's absorption features are used. Of the planet's absorption features, CH$_4$ lines contribute the most signal to the detection. CO$_2$ which has more spectral lines in this wavelength range is not well detected in $50$ hours. This is because CO$_2$ has strong aliases in its auto-correlation function in this wavelength regime meaning the telluric residuals create more noise in the CCF than they do for CH$_4$, hindering the detection. An improved data reduction might result in a stronger detection for CO$_2$. For the stellar spectrum only model no significant detection is made in $50$ hours. Again, this is due to the imperfect data reduction which leaves behind a small residual stellar spectrum in the data. The residual strongly correlates with the stellar spectrum only model creating correlated noise in the CCF. The signal-to-noise of the residuals grows faster than signal-to-noise of the planet which prevents the detection of the planet's signal.

%%%%%%%%%%%%%%%%%%%%%%%%%%%%%%%%%%%%%%%%%%%%%%%%%%
\subsection{Time Available for Observations}
\label{sec:timeavalible}

\begin{figure*}
	\includegraphics[width=\textwidth]{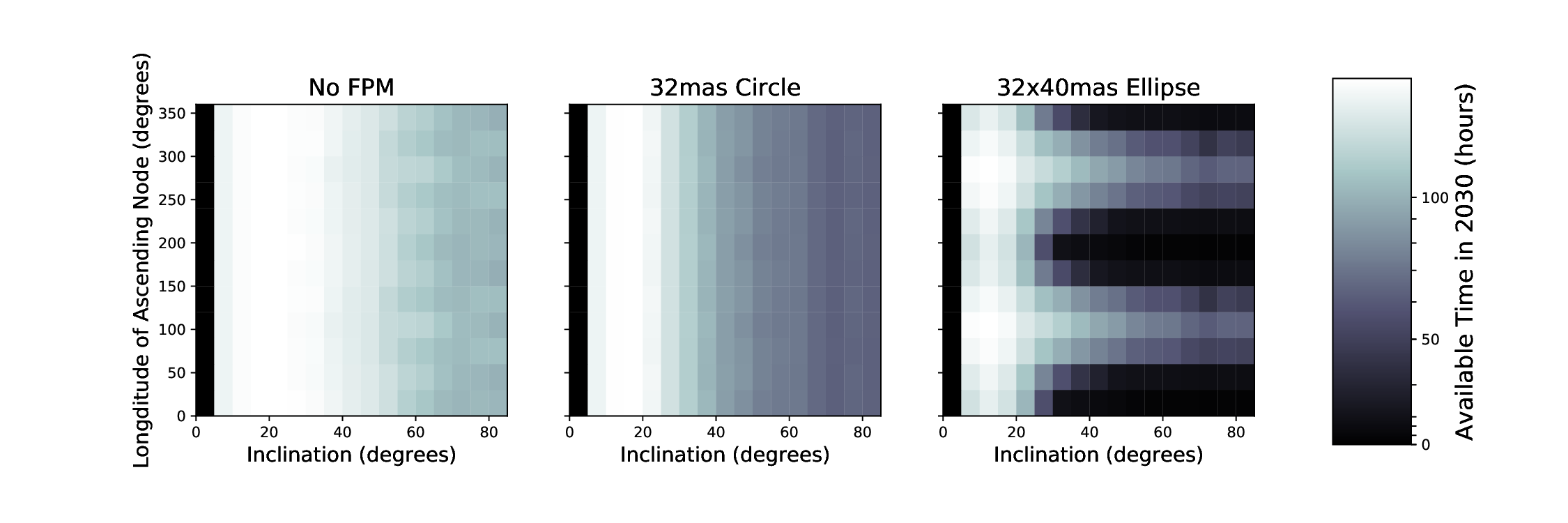}
    \caption{The amount of time available for which our observing conditions for Proxima b are met between 2030 January 01 and 2031 January 01 as a function of the inclination and longitude of the ascending node of Proxima b's orbit. Each plot shows the results for the different FPMs.}
    \label{fig:observability}
\end{figure*}

The amount of Proxima b's orbit covered by each mask and therefore the amount of time it could be observed depends on the currently unknown shape and orientation of the orbit. We calculate the observation time available for different orbits between midday on the $1^{st}$ January 2030 and midday on the $1^{st}$ January 2031 for the observing criteria described in Section \ref{sec:scheduler} with the additional conditions: i) for `no FPM', Proxima b's is not on a spaxel which is above the persistence limit, and ii) for the circular and elliptical FPMs, there are no spaxels above the persistence limit assuming an integration time of $\SI{60}{\second}$. The result of these calculations is illustrated in Fig. \ref{fig:observability} where each plot shows time available for a range of inclinations and longitude of the ascending node of the orbit. For almost all orbital inclinations with the no FPM and circular FPM, there is at least $70$ hours of time available to observe Proxima b. For the elliptical mask, there is less time available for the orbits that more frequently align with the elongated direction of the mask. We do not calculate the amount of time available for the offset mask case as it will depend on the maximum allowable offset which is currently unknown.

%%%%%%%%%%%%%%%%%%%%%%%%%%%%%%%%%%%%%%%%%%%%%%%%%%
%%%%%%%%%%%%%%%%%%%%%%%%%%%%%%%%%%%%%%%%%%%%%%%%%%
%%%%%%%%%%%%%%%%%%%%%%%%%%%%%%%%%%%%%%%%%%%%%%%%%%
\section{Discussion}
\label{sec:discussion}
%%%%%%%%%%%%%%%%%%%%%%%%%%%%%%%%%%%%%%%%%%%%%%%%%%
\subsection{What if Proxima b's orbit is not well constrained?}
\label{sec:unknownorbit}
It is impossible to constrain Proxima b's orbit with radial velocities alone. GAIA's astrometric precision makes it sensitive only to near face-on orbits but its observing cadance may not be sufficient to measure the astrometric wobble in Proxima Centauri's position induced by Proxima b. For Proxima b's orbit to be known before these observations, it would have to be observed by another integral field or long-slit spectrograph as detection via direct imaging alone without cross-correlation is unlikely with current planned instruments for ELT. If the orbit of Proxima b is not well known then observations could be made when Proxima b is at quadrature, the time of which is known from radial velocity measurements. At quadrature, Proxima b's separation from Proxima Centauri will be at its maximum, it will have a different Doppler shift to Proxima Centauri (if the orbit is not too close to face-on), and approximately half of its surface should be illuminated thus meeting our observing criteria from Section \ref{sec:scheduler}. Therefore, by observing within a small phase range around one of the quadrature points -- approximately the same point in its orbit for each observation -- the observations will only need to be aligned to celestial coordinates and Doppler shifted to remove the barycentric motion in order to align the planets signal for the analysis in Section \ref{sec:crosscorrelation}. However, this technique may require more observations depending on whether the mask blocks the planet at quadrature. 

\subsection{Feasibility of the observations}
Assuming only observations with the simulated seeing ($\SI{0.57}{\arcsecond}$) or better are made then around $30$ per cent\footnote{\url{https://www.eso.org/sci/facilities/eelt/docs/ESO-193696_2_Observatory_Top_Level_Requirements.pdf}} of the time available (see Fig. \ref{fig:observability}) can be used. Worse seeing could be used but these would contribute less to the detection of the exoplanet due to increased stellar contamination at the exoplanet's location. 

With the circular mask it is possible, in most cases, to obtain $20$ hours of integration time under the right conditions within $\sim1$ year with the exception of near face-on orbits. For the elliptical mask it may take longer depending on the alignment of Proxima b's orbit. Near face-on orbits have very little time available due to the velocity criterion however, the inclinations where this effect is significant correspond to over a factor of $10$ difference between the measured minimum mass and the true mass of Proxima b which changes the amount of reflected light and therefore time required significantly. It should be noted that this will not take up a large fraction of the telescope's available time, merely that it will require a series of observations spread out over the course a few years due to the strict restrictions on when the observations can take place.

%%%%%%%%%%%%%%%%%%%%%%%%%%%%%%%%%%%%%%%%%%%%%%%%%%
\subsection{Caveats of the simulation}
\label{sec:caveats}
Simulations rarely capture all of the nuances of real observations and the simulations presented here have a number of caveats that will affect our results. Firstly, we use the same simplified PSF with no wavelength dependence for each observation. This results in the speckles behaving as white noise sources instead of 1/f noise. This simplification decreases the amount of time required for a detection. Secondly, we do not include scattering caused by the sharp edge of the focal plane mask. Although this effect is not expected to be significant, increased light leakage around the mask may increase the amount of integration time required. Since Proxima b can only be observed close to the mask edge, it will be pushing the limits of what HARMONI can do. Proof will ultimately be realised on sky, and greatly aided by a known orbit for Proxima b.

%%%%%%%%%%%%%%%%%%%%%%%%%%%%%%%%%%%%%%%%%%%%%%%%%%
\subsection{Selecting a mask}
\label{sec:selectmask}
%%%%%%%%%%%%%%%%%%%%%%%%%%%%%%%%%%%%%%%%%%%%%%%%%%
\subsubsection{Observing Proxima b with no FPM}
\label{sec:nochange}
Fig \ref{fig:result} indicates that observing without a FPM would be ideal for characterising Proxima b, however, our simulations do not include a realistic treatment of the effects of persistence in the detector. In reality, it will not be possible to use $\SI{60}{\second}$ integrations without a FPM as the star will cause persistence that may severely impact these and subsequent observations, particularly of faint targets. To observe without a FPM, the integration time would have to be reduced to $\sim\SI{0.5}{\second}$ resulting in a poor duty cycle and drastically increased the read noise which would strongly hinder the retrieval of the exoplanet's spectrum.

%%%%%%%%%%%%%%%%%%%%%%%%%%%%%%%%%%%%%%%%%%%%%%%%%%
\subsubsection{Using the Circular FPM with radius $\SI{32}{\mas}$}
\label{sec:circularfpm}
The next best performing mask as indicated by Fig. \ref{fig:result} is the Circular FPM with radius $\SI{32}{\mas}$. However as seen in Fig. \ref{fig:leakage}, of the remaining masks, this has the largest amount of light leaking around its edge which may impact the detection. Additionally, to avoid persitance with $\SI{60}{\second}$ integrations, the observations are limited to low airmass.

%%%%%%%%%%%%%%%%%%%%%%%%%%%%%%%%%%%%%%%%%%%%%%%%%%
\subsubsection{Observing Proxima b by offsetting the currently planned FPM}
\label{sec:offsetfpm}
Offsetting the mask does not result in as strong a detection as the no FPM and circular FPM cases. Additionally, if the orbit of Proxima b is not well constrained as it will not be possible to predict what offset to use to detect the exoplanet, even if observations are made at quadrature. This increases the amount of telescope time needed for the offset mask as trial and error would be required to find the right offset drastically lowering the efficiency of these observations.

%%%%%%%%%%%%%%%%%%%%%%%%%%%%%%%%%%%%%%%%%%%%%%%%%%
\subsubsection{Using the Elliptical FPM with dimensions $\SI{32}{}\times\SI{40}{\mas}$}
\label{sec:ellipticalfpm}
The elliptical FPM does not perform as well as the other masks as the planet is on average closer to its edge however, given the caveats discussed this is likely the most viable option. Nonetheless, there is still leakage around the mask edge and, due to its elongation, the mask could cover the orbit of Proxima b even at quadrature. If Proxima b's orbit it unknown, this would lower the efficiency as the planet could be behind the mask during some observations.

%%%%%%%%%%%%%%%%%%%%%%%%%%%%%%%%%%%%%%%%%%%%%%%%%%
\subsubsection{Changing the apodizer}
\label{sec:changeapodizer}
Reducing the light leakage around the mask edge would likely benefit these observations. While not studied in this work, one way to potentially achieve this is to change the apodizer to decrease the size of the central core of the PSF. This would allow the FPM to be decreased in size, whilst still protecting the detector from saturation and persistence, and making the observations more robust against pointing errors. However, as with changing the FPM, changing the apodizer requires careful consideration of its effect on other science cases.

%%%%%%%%%%%%%%%%%%%%%%%%%%%%%%%%%%%%%%%%%%%%%%%%%%
\subsection{Other Instruments}
\label{sec:otherins}
HARMONI is not the only instrument that could be used to observe this target. Section \ref{sec:intro} lists a number of instruments that might also be able to detect the atmosphere of Proxima b. However, HARMONI will be one of the first instruments available and therefore one of the first that could make these observations. Other instruments that could potentially detect the atmosphere of Proxima b at around the same time are METIS@ELT \citep{METIS}, GMagAO-X@GMT \citep{GMagAOX} and IRIS@TMT \citep{IRIS}. Of these, GMagAO-X and IRIS will both be sensitive to the reflected light of Proxima b like HARMONI however owing to smaller primary mirrors, they have lower spatial resolutions. Careful consideration would be needed to determine if these instruments could observe and characterise Proxima b. METIS will be sensitive to the thermal emission ($\SI{3}{}$--$\SI{5}{\micro\meter}$) of Proxima b and has a higher spectral resolving power than HARMONI ($R=100,000$), although its spaxel scale is larger \citep[$\SI{8.2}{\mas} \times \SI{21}{\mas}$; ][]{Brandl2021}, Proxima b is resolvable with this instrument ($\lambda/D$ at $\SI{4}{\micro\meter}$ is $\SI{21}{\mas}$). 

With the potential of all these instruments to observe Proxima b, we have the opportunity to obtain one of the most detailed spectra of an Earth-like exoplanet to date, spanning both the reflection and thermal emission of this world. Although one could argue that each instrument could be sensitive to similar atmospheric properties, given this planet could potentially host a habitat similar to Earth, the extra degree of certainty and complementary overlap would be ideal. Additionally, having access to both the thermal emission and reflection of this planet may give us additional information on its atmospheric composition, temperature distribution, cloud and haze properties, and energy balance \citep[e.g.][]{Crossfield2013,Morley2015,Steinrueck2023}, for a full, holistic study.

\subsection{Other Temperate Terrestrial Exoplanets}
\label{sec:othertargets}
We have only considered the temperate terrestrial exoplanet Proxima b in this work as, of the known exoplanets of this type, it is the easiest to observe in reflected light. The next best known temperate exoplanet to observe is Wolf 1061 c which has an on sky separation, also inside the IWA, of approximately $\SI{20}{\mas}$ (although this exoplanet is close to the boundary between rocky and gas giant exoplanets so it may not be terrestrial). This exoplanet's separation is too small to observe with HARMONI unless the apodizer is changed. It is possible that there are temperate terrestrial exoplanets orbiting nearby stars with larger on sky separation that we have yet to identify. However, for an Earth-like planet to be bright enough compared to its star for current instrumentation to detect, it must orbit close to the star which limits suitable targets to nearby M-dwarfs. Proxima Centauri is the closest M-dwarf to the solar system and as such Proxima b is almost certainly the best target we will have for these observations.

%%%%%%%%%%%%%%%%%%%%%%%%%%%%%%%%%%%%%%%%%%%%%%%%%%
%%%%%%%%%%%%%%%%%%%%%%%%%%%%%%%%%%%%%%%%%%%%%%%%%%
%%%%%%%%%%%%%%%%%%%%%%%%%%%%%%%%%%%%%%%%%%%%%%%%%%
\section{Conclusions}
\label{sec:conclusions}

We simulate observations made with the HCAO mode of HARMONI@ELT to determine the viability of using the molecule mapping technique to characterise the atmosphere of the terrestrial exoplanet Proxima b in reflected light. HARMONI's HCAO mode uses an apodizer to suppress the diffracted star light in a ring around the star and a focal plane mask to suppress the central diffraction peak. If Proxima Centauri is observed on axis using the apodizer with the smallest inner working angle and the smallest FPM, Proxima b's orbit will be fully obscured by the FPM.

Here we explored relatively minor modifications to the current design of HARMONI to counteract this. The design change required depends on how well the orbit of Proxima b is constrained and how large of an impact the change will have on other science cases. The instrument's design will likely have to be committed to hardware before more information on the planet's orbit is available. If Proxima b's orbit will be well known at the time of the observations, then no changes may be required as the star could be offset from the center of the field of view allowing the exoplanet to be unobscured by the FPM. Our simulations show a detection could be possible with this set up but do not account for the increased stray light due to the star being close to the edge of the mask. If the orbit will not be well known then the current FPM could be replaced by a smaller one. This work indicates that with our caveats, doing so should allow characterisation of Proxima b, requiring at least $20$ hours and ideally at least $30$ hours of integration time for a $S/N\geq5$ detection assuming an orbital inclination of $\SI{45}{\degree}$. The masks investigated in this work typically limit observations to $2$ hours per night so around $10$ such observations would be needed which, for inclination of $\SI{45}{\degree}$, could be obtained over a period as short as $4$ months. The signal-to-noise of this detection is dominated by the atmospheric features in the planet's spectrum and is particularly sensitive to the biosignature CH$_4$. This is highlighted when the star-only template is used as the cross-correlation model. In this case there is no significant peak in the cross-correlation coefficients at the expected position and velocity of the planet. This is due to correlated noise from the residual stellar contamination which creates signals in the cross-correlation stronger than the planet's signal inhibiting the detection. Finally, while not investigated here, changing both the FPM and modifying the apodizer to have a smaller central core, would likely improve the detection of Proxima b with HARMONI. This would help mitigate potential issues with stray light and may allow other temperate terrestrial exoplanets to be characterised depending on the new inner working angle.

As changing the mask will require removing one of the current masks, and changing the apodizer would change the range of separations the HCAO mode can be used for, careful consideration would be required as to how these changes effect other science cases.

%%%%%%%%%%%%%%%%%%%%%%%%%%%%%%%%%%%%%%%%%%%%%%%%%%
%%%%%%%%%%%%%%%%%%%%%%%%%%%%%%%%%%%%%%%%%%%%%%%%%%
%%%%%%%%%%%%%%%%%%%%%%%%%%%%%%%%%%%%%%%%%%%%%%%%%%
\section*{Acknowledgements}

We thank the anonymous referee for their helpful comments that improved the quality of the manuscript.

SRV and JLB acknowledge funding from the European Research Council (ERC) under the European Union’s Horizon 2020 research and innovation program under grant agreement No 805445.
MPS acknowledges funding support from the Ram\'on y Cajal program of the Spanish Ministerio de Ciencia e Innovaci\'on (RYC2021-033094-I).
MH and AV acknowledge funding from the European Research Council (ERC) under the European Union’s Horizon 2020 research and innovation program under grant agreement No 757561.
NT and FC acknowledge support from the Science and Technology Facilities Council (UKRI) grants ST/X002322/1 and ST/S001409/1.
%RTP acknowledge support from the European Research Council Advanced grant EXOCONDENSE, No 740963.

This research has made use of the NASA Exoplanet Archive, which is operated by the California Institute of Technology, under contract with the National Aeronautics and Space Administration under the Exoplanet Exploration Program. This research has made use of NASA's Astrophysics Data System Bibliographic Services and the SIMBAD database, operated at CDS, Strasbourg, France. 

%%%%%%%%%%%%%%%%%%%%%%%%%%%%%%%%%%%%%%%%%%%%%%%%%%
\section*{Data Availability}

The data underlying this article will be shared on reasonable request to the corresponding author.
This work has made use of \textsf{numpy} \citep{NumPy2020}, \textsf{scipy} \citep{scipy_2020}, \textsf{matplotlib} \citep{matplotlib2007}, and \textsf{astropy},\footnote{\url{https://www.astropy.org}} a community-developed core Python package and an ecosystem of tools and resources for astronomy \citep{astropy:2013, astropy:2018, astropy:2022}.



\bibliographystyle{mnras}
\bibliography{example, proxima, HARMONI, Molecule_Mapping, misc, HRS, ELT_inst, python_packages} 

%%%%%%%%%%%%%%%%%%%%%%%%%%%%%%%%%%%%%%%%%%%%%%%%%%

% Don't change these lines
\bsp	% typesetting comment
\label{lastpage}
\end{document}